\documentclass[twocolumn,twosided]{IEEEtran}
\usepackage{amsmath,amstext,amssymb,epsf}

 \newtheorem{lemma}{Lemma}[section]
 \newtheorem{theorem}[lemma]{Theorem}

 \newtheorem{definition}[lemma]{Definition}
\newtheorem{rem}[lemma]{Remark}
\newenvironment{remark}{\begin{rem}}{\hspace*{\fill}$\diamondsuit$\end{rem}}
 \newtheorem{ex}[lemma]{Example}
\newenvironment{example}{\begin{ex}}{\hspace*{\fill}$\diamondsuit$\end{ex}}
                                                                                
\numberwithin{equation}{section} 

\newtheorem{notation}[lemma]{Notation}

\newcommand{\lea}{\stackrel{{}_+}{<}}
\newcommand{\gea}{\stackrel{{}_+}{>}}
\newcommand{\eqa}{\stackrel{{}_+}{=}}

\newcommand{\eps}{\epsilon}

\newcommand{\len}[2]{l_{#1}(#2)}

\newcommand{\soph}{\mbox{\rm soph}}

\begin{document}
\title{Meaningful Information}
\author{Paul Vit\'anyi\thanks{
Manuscript received xxx, 2002;
revised yyy 2004. The material of this paper was presented in part in
{\em 13th International Symposium on Algorithms and Computation 
Vancouver, Canada, November 20-23, 2002}.
This work supported in part
by the EU fifth framework project QAIP, IST--1999--11234,
the NoE QUIPROCONE IST--1999--29064,
the ESF QiT Programmme, and the EU Fourth Framework BRA
NeuroCOLT II Working Group
EP 27150, and the EU NoE PASCAL.
Address: CWI, Kruislaan 413,
1098 SJ Amsterdam, The Netherlands.
Email: {\tt Paul.Vit\'anyi@cwi.nl}.}
}

\markboth{IEEE Transactions on Information Theory, VOL. XX, NO Y, MONTH 2004}{P.M.B. Vit\'anyi: Meaningful Information}

\maketitle

\begin{abstract}
The information in an individual finite object (like a binary string)
is commonly measured by its Kolmogorov complexity. One can divide
that information into two parts: the information accounting  
for the useful regularity present in the object and the information
accounting for the remaining accidental information. There can be several ways
(model classes) in which the regularity is expressed. Kolmogorov
has proposed the model class of finite sets, generalized later
to computable probability mass functions. The resulting theory,
known as Algorithmic Statistics, analyzes the algorithmic
sufficient statistic when the statistic is restricted to the given
model class. However, the most
general way to proceed is perhaps to express the
useful information  as a recursive function.  
The resulting measure has been called the ``sophistication'' of the object.
We develop the theory of recursive
functions statistic, the maximum and minimum value,
the existence of absolutely nonstochastic objects (that have
maximal sophistication---all the information in them is meaningful
and there is no residual randomness), 
determine its relation with the more restricted model
classes of finite sets, and computable probability distributions,
in particular with respect to the algorithmic (Kolmogorov)
minimal sufficient statistic, the relation to the halting problem
and further algorithmic properties.


\end{abstract}

\section{Introduction}
The information contained by an individual
finite object (like a finite binary string) is objectively measured
by its Kolmogorov complexity---the length of the shortest binary program
that computes the object. Such a shortest program contains no redundancy:
every bit is information; but is it meaningful information? 
If we flip a fair coin to obtain a finite binary string, then with overwhelming
probability that string constitutes its own shortest program. However,
also with overwhelming probability all the bits in the string are meaningless
information, random noise. On the other hand, let an object
$x$ be a sequence of observations of heavenly bodies. Then $x$
can be described by the binary
string $pd$, where $p$ is the description of 
the laws of gravity, and the observational
parameter setting, while $d$ is the data-to-model code accounting
for the (presumably Gaussian) measurement error in the data.
This way we can divide the information in $x$ into
meaningful information $p$ and data-to-model information $d$. 

The main task for statistical inference and learning theory is to
distil the meaningful information present in the data. The question
arises whether it is possible to separate meaningful 
information from accidental information, and if so, how.

In statistical theory, every function of the data
is called a ``statistic'' of the data. 
The central notion in probabilistic statistics is that of
a ``sufficient'' statistic, introduced
by the father of statistics R.A. Fisher~\cite{Fi22}:
       ``The statistic chosen should summarise the whole of the relevant
information supplied by the sample. This may be called
       the Criterion of Sufficiency $\ldots$
In the case of the normal curve
of distribution it is evident that the second moment is a
       sufficient statistic for estimating the standard deviation.''
For traditional problems, dealing with frequencies over small
sample spaces, this approach is appropriate. But for
current novel applications, average relations are often
irrelevant, since the part of the support of the probability
density function that will ever be observed has about zero
measure. This is the case in, for example, complex video and sound
analysis.
There arises the problem that for individual cases the selection
performance may be bad although the performance is good on average.
There is also the problem of what probability means,
whether it is subjective, objective, or exists at all.

To simplify matters,
and because all discrete data
can be binary coded, we consider only data samples
that are finite binary strings.
The basic idea is to
 found statistical theory on finite combinatorial
principles independent of probabilistic assumptions, as the relation
between the individual data and its explanation (model).
We study extraction of meaningful information
in an initially limited setting where
this information be represented by a finite set (a model) of which the
object (the data sample) is a typical member. Using the theory
of Kolmogorov complexity, we can rigorously express and quantify
typicality of individual objects.
But typicality in itself is not necessarily a significant property:
every object is typical in the singleton set containing only
that object. More important is the following Kolmogorov complexity
analog of probabilistic minimal sufficient statistic which implies typicality:
The two-part description of the smallest finite
set, together with the index of the object in that set, is as concise
as the shortest one-part description of the object.
The finite set 
models the regularity
present in the object (since it is a typical element of the set). This
approach has been generalized to 
computable probability mass functions.
The combined theory has been developed in detail in \cite{GTV} and
called ``Algorithmic Statistics.''

Here we study the most general form of algorithmic statistic:
recursive function models. 
In this setting the issue of meaningful
information versus accidental information is put in its starkest form;
and in fact, has been around for a long time in various imprecise forms
unconnected with the sufficient statistic approach:
The issue has sparked
the imagination and entered scientific
popularization in \cite{GM94} as ``effective complexity''
(here ``effective'' is apparently used in the sense of
``producing an effect'' rather than ``constructive''
as is customary in the theory of computation). It is time that it receives
formal treatment.
Formally, we study the minimal length of a total recursive function
that leads to an optimal length two-part code of the object being described.
(``Total'' means the function value is defined for all arguments in the
domain,
and ``partial'' means that the function is possibly not total.)
This minimal length has been called the ``sophistication'' of the object
in \cite{Ko87,Ko88} in a different, but related,
setting of compression
and prediction properties of infinite sequences.
That treatment is
technically sufficiently vague so as to have
no issue for the present work. 
We develop the notion based on prefix
Turing machines, rather than on a variety of monotonic Turing machines
as in the cited papers. Below we 
describe related work in detail and
summarize our results. 
Subsequently, we formulate our problem in the formal
setting of computable two-part codes.

\subsection{Related Work}
A.N. Kolmogorov in 1974 \cite{Ko74} proposed
an approach to a non-probabilistic statistics
based on Kolmogorov complexity. An essential feature of this approach
is to separate the data into meaningful information (a model)
and meaningless information (noise). 
Cover~\cite{Co85,CT91} attached the name
``sufficient statistic'' to a 
model of which the data is a ``typical''
member.
In Kolmogorov's initial setting 
the models are finite sets. As Kolmogorov himself pointed
out, this is no real restriction:
the finite sets model class is equivalent, up to a logarithmic additive
term, to the model class of computable probability density functions, as
studied in \cite{Sh83,Sh99,ViLi99}.
Related aspects of ``randomness deficiency'' 
were formulated in~\cite{Ko83,KU88} and
studied in~\cite{Sh83,Vy87}.
Despite its evident epistemological
prominence in the theory of hypothesis selection
and prediction, only selected
aspects of the theory
were studied in these references.
Recent work \cite{GTV} can be considered as a comprehensive investigation 
into the sufficient statistic for finite set models and computable
probability density function models.
Here we extend the approach to
the most general form: the model class of
total recursive functions.
This idea was pioneered by \cite{Ko87,Ko88}
who, unaware of a statistic connection, coined the cute word ``sophistication.''
The algorithmic (minimal) sufficient statistic 
was related to an applied form in \cite{ViLi99,GLV00}: the well-known
``minimum description length'' principle \cite{BRY} in statistics
and inductive reasoning. 

In another paper \cite{VeVi02}
(chronologically following the present paper) we
comprehensively treated all stochastic properties of the data in terms of
Kolmogorov's so-called structure functions. The sufficient statistic
aspect, studied here, covers only part of these properties.
The results on the structure functions, including (non)computability
properties, are valid, 
up to logarithmic
additive terms, also for
the model class of total recursive functions, as studied here.

\subsection{This Work:}
It will be helpful for the reader to be familiar with
initial parts of \cite{GTV}. In \cite{Ko74}, Kolmogorov observed 
that randomness of an object in the sense of 
having high Kolmogorov complexity is being random in 
just a ``negative'' sense.
That being said, 
we define the notion of sophistication (minimal sufficient statistic
in the total recursive function model class).
It is demonstrated to be meaningful (existence and
nontriviality). We then establish lower and upper bounds
on the sophistication, and we show that there are
objects the sophistication
achieves the upper bound. In fact, these are objects in which all information is
meaningful and there is (almost) no accidental information. 
That is, the simplest explanation of such an object
is the object itself. In the simpler setting of finite set statistic
the analogous objects were called ``absolutely non-stochastic''
by Kolmogorov. If such objects have high Kolmogorov complexity, 
then they can only
be a random outcome of a ``complex'' random process,
and Kolmogorov questioned whether such random objects, being random in just
this ``negative'' sense, can occur in nature.
But there are also objects that are random in the sense of having
high Kolmogorov complexity, but simultaneously are
are typical outcomes of ``simple'' random processes. These
were therefore said to be  random in a ``positive'' sense \cite{Ko74}.
An example are the  strings of maximal
Kolmogorov complexity; those are very 
unsophisticated (with sophistication
about 0), and are typical outcomes of tosses with a fair coin---a very
simple random process.
We subsequently establish the equivalence between
sophistication and the algorithmic minimal sufficient statistics
of the finite set class and the probability mass function class.
Finally, we investigate the algorithmic properties of sophistication:
nonrecursiveness, upper semicomputability, and intercomputability
relations of Kolmogorov complexity, sophistication, halting sequence.

\section{Preliminaries}

A {\em string} is a finite binary sequence, an element of $\{0,1\}^*$. 
If $x$ is a string then the {\em length} $l(x)$ denotes the number
of bits in $x$.
We identify
${\cal N}$, the natural numbers, and $\{0,1\}^*$ according to the
correspondence
\[(0, \epsilon ), (1,0), (2,1), (3,00), (4,01), \ldots \]
Here $\epsilon$ denotes the {\em empty word}.
Thus, $l(\epsilon)=0$.
The emphasis is on binary sequences only for convenience;
observations in any alphabet can be so encoded in a way
that is `theory neutral'.
Below we will use the natural numbers and the strings
interchangeably.

A string $y$
is a {\em proper prefix} of a string $x$
if we can write $x=yz$ for $z \neq \epsilon$.
 A set $\{x,y, \ldots \} \subseteq \{0,1\}^*$
is {\em prefix-free} if for any pair of distinct
elements in the set neither is a proper prefix of the other.
A prefix-free set is also called a {\em prefix code} and its
elements are called {\em code words}.
An example of a
prefix code, that is useful later,
encodes the source word $x=x_1 x_2 \ldots x_n$ by the code word
\[ \overline{x} = 1^n 0 x .\]
This prefix-free code
is called {\em self-delimiting}, because there is fixed computer program
associated with this code that can determine where the
code word $\bar x$ ends by reading it from left to right without
backing up. This way a composite code message can be parsed
in its constituent code words in one pass, by the computer program.
(This desirable property holds for every prefix-free
encoding of a finite set of source words, but not for every
prefix-free encoding of an infinite set of source words. For a single
finite computer program to be able to parse a code message the encoding needs
to have a certain uniformity property like the $\overline{x}$ code.)
Since we use the natural numbers and the strings interchangeably,
$l(\bar x)$ where $x$ is ostensibly an integer, means the length
in bits of the self-delimiting code of the string with index $x$.
On the other hand, $\overline{l(x)}$ where $x$ is ostensibly a 
string, means the self-delimiting code of the string
with index the length $l(x)$ of $x$.
Using this code we define
the standard self-delimiting code for $x$ to be
$x'=\overline{l(x)}x$. It is easy to check that
$l(\overline{x} ) = 2 n+1$ and $l(x')=n+2 \log n +1$.
Let $\langle \cdot \rangle$ denote a standard invertible
effective one-one encoding from ${\cal N} \times {\cal N}$
to a subset of ${\cal N}$.
For example, we can set $\langle x,y \rangle = x'y$
or $\langle x,y \rangle = \bar xy$.
We can iterate this process to define
$\langle x , \langle y,z \rangle \rangle$,
and so on.

\subsection{Kolmogorov Complexity}
For definitions, notation, and an
introduction to Kolmogorov complexity, see \cite{LiVi97}.
Informally, the Kolmogorov complexity, or algorithmic entropy, $K(x)$ of a
string $x$ is the length (number of bits) of a shortest binary
program (string) to compute
$x$ on a fixed reference universal computer
(such as a particular universal Turing machine).
Intuitively, $K(x)$ represents the minimal amount of information
required to generate $x$ by any effective process.
The conditional Kolmogorov complexity $K(x | y)$ of $x$ relative to
$y$ is defined similarly as the length of a shortest program
to compute $x$, if $y$ is furnished as an auxiliary input to the
computation.
For technical reasons we use a variant of complexity,
so-called prefix complexity, which is associated with Turing machines
for which the set of programs resulting in a halting computation
is prefix free.
We realize prefix complexity by considering a special type of Turing
machine with a one-way input tape, a separate work tape,
and a one-way output tape. Such Turing
machines are called {\em prefix} Turing machines. If a machine $T$ halts
with output $x$
after having scanned all of $p$ on the input tape,
but not further, then $T(p)=x$ and
we call $p$ a {\em program} for $T$.
It is easy to see that
$\{p : T(p)=x, x \in \{0,1\}^*\}$ is a {\em prefix code}.
\begin{definition}
\rm
A function
$f$ from the natural numbers to the natural numbers is {\em partial recursive},
or {\em computable},
if there is a Turing machine $T$ that computes it: $f(x)=T(x)$ for all $x$
for which  either $f$ or $T$ (and hence both) are defined. 
This definition can be extended to (multi-tuples of) 
rational arguments and values.
\end{definition}
Let $T_1 ,T_2 , \ldots$ be a standard enumeration
of all prefix Turing machines with a binary input tape,
for example the lexicographical length-increasing ordered syntactic
prefix Turing machine descriptions, \cite{LiVi97},
and let $\phi_1 , \phi_2 , \ldots$
be the enumeration of corresponding functions
that are computed by the respective Turing machines
($T_i$ computes $\phi_i$).
These functions are the partial recursive functions
of effectively prefix-free encoded
arguments. The Kolmogorov complexity
of $x$ is the length of the shortest binary program
from which $x$ is computed by such a function.
\begin{definition}
\rm
The {\em prefix Kolmogorov complexity} of $x$ is
                  \begin{equation}\label{eq.KC}
K(x) = \min_{p,i}  \{l(\bar i) +  l(p):
T_i (p)=x \} ,
                  \end{equation}
where the minimum is taken over $p \in \{0,1\}^*$ and  $i
\in \{1,2, \ldots \}$.
For the development of the theory we
actually require
the Turing machines to use {\em auxiliary} (also
called {\em conditional})
information, by equipping the machine with a special
read-only auxiliary tape containing this information at the outset.
Then, the {\em conditional version} $K(x \mid y)$ of the prefix
Kolmogorov complexity of $x$
given $y$ (as
auxiliary information) is
is defined similarly as before,
and the unconditional version is set to  $K(x)=K(x  \mid \epsilon)$.
\end{definition}

\begin{notation}
\rm
  From now on, we will denote by $\lea$ an inequality to within an
additive constant, and by $\eqa$ the situation when both $\lea$ and
$\gea$ hold.
\end{notation}

\subsection{Two-Part Codes}
Let $T_1,T_2, \ldots$ be the standard
enumeration of Turing machines, and let 
$U$ be a standard Universal Turing machine
satisfying $U(\langle i, p \rangle) = T_i(p)$ for all indices $i$
and programs $p$. We fix $U$ once and for all and call it the {\em reference
universal prefix Turing machine}. 
The shortest program to compute $x$ by $U$ is
denoted as $x^*$  (if there is more than one
of them, then $x^*$ is the first one in standard enumeration).
It is a deep and useful fact
 that the shortest effective description of an object $x$
can be expressed in terms of a {\em two-part code}:
the first part describing an appropriate Turing machine
and the second part describing
the program that interpreted by the Turing machine
reconstructs $x$. The essence of the theory
is the Invariance Theorem, that can be informally stated as follows:
For convenience, in the sequel
 we simplify notation
and write $U(x,y)$ for $U(\langle x, y\rangle)$. 
Rewrite
             \begin{align*}
K(x) & \min_{p,i} \{l(\bar i) + l(p):T_i(p) =x\}
\\&  \nonumber
\min_{p,i} \{2l(i)+l(p)+1:T_i(p) =x\}
\\&  \leq
\nonumber
\min_{q} \{l(q): U( \epsilon, q ) =x\}+2l(u)+1
\\&  \leq
\nonumber
\min_{r,j} \{l(j^*)+l(r): U(\epsilon, j^* \alpha r )
=T_j(r) =x\}
\\&
\nonumber
\hspace{2.5in} +2l(u)+1
\\&  \lea
K (x).
\nonumber
\end{align*}
Here the minima are taken over
$p,q,r \in \{0,1\}^*$ and $i,j \in \{1,2, \ldots\}$.
The last equalities are obtained by
using the universality of 
$U=T_u$ with $l(u) \eqa 0$.
As consequence,
\begin{align*}
& K (x) \eqa \min_{r,j} \{l(j^*)+l(r): U(\epsilon, j^* \alpha r ) 
=T_j(r) =x\}
\\&K(x) \eqa K_U(x) = \min \{l(q): U(\epsilon , q) = x \}.
\end{align*}
Thus, $K(x)$ and $K_U(x)$ differ by at most an additive constant 
depending on the choice of 
$U$. It is standard to use
\begin{equation}\label{eq.KCU}
K(x) \equiv K_U(x)
\end{equation}
instead of \eqref{eq.KC}
as the
definition of {\em prefix Kolmogorov complexity}, \cite{LiVi97}.
However, we highlighted definition \eqref{eq.KC}
to bring out the two-part code nature.
By universal logical principles, the resulting theory is recursively 
invariant under adopting either definition \eqref{eq.KC} or 
definition \eqref{eq.KCU},
as long as we stick to one choice.
If $T$ stands for a literal
description of the prefix Turing machine $T$ in standard format,
for example the index $j$ when $T=T_j$, then we can write $K(T) \eqa K(j)$.
The string $j^*$ is a shortest self-delimiting
program of $K(j)$ bits from which $U$ can compute $j$,
and subsequent execution of the next self-delimiting fixed program
$\alpha$ will compute $\bar j$ from $j$.  Altogether,
this has the effect that $U( \epsilon, j^* \alpha r )
=T_j(r)$.
If $(j_0,r_0)$ minimizes the expression above, then $T_{j_0}(r_0)=x$,
and hence $K(x|j_0) \lea l(r_0)$,
and $K(j_0)+l(r_0) \eqa K(x)$. 
It is straightforward that $K(j_0)+K(x| j_0) \gea K(x,j_0)
\gea K(x)$, and therefore we have $l(r_0) \lea K(x|j_0)$.
Altogether, $l(r_0)\eqa K(x|j_0)$.
Replacing the minimizing $j=j_0$ by the minimizing
$T=T_{j_0}$ and $l(r_0)$ by $K(x|T)$, we can
rewrite the last displayed equation as
\begin{equation}\label{eq.kcmdl}
K(x) \eqa \min_T \{K(T)+ K(x \mid T): T \in \{T_0,T_1, \ldots \}\}.
\end{equation}

\subsection{Meaningful Information}
Expression \eqref{eq.kcmdl} emphasizes the two-part code 
nature of Kolmogorov complexity:
using the regular aspects of $x$ to maximally compress.
Suppose we consider an ongoing time-series $0101 \ldots$ and we
randomly stop gathering data after having obtained the initial segment 
\[ x = 10101010101010101010101010.
\]
We can encode this $x$ by a small Turing machine representing 
``the repeating pattern is 01,'' and which computes
$x$, for example,  from the program ``13.''
Intuitively, the Turing machine part of the code
squeezes out the {\em regularities} in $x$. What
is left are irregularities, or {\em random aspects}
of $x$ relative to that Turing machine. The minimal-length
two-part code squeezes out regularity only insofar as
the reduction in the length of the description of random aspects
is greater than the increase in the regularity description.
In this setup the number of repetitions of the significant pattern
is viewed as the random part of the data. 

This interpretation of $K(x)$ as the shortest length of
a two-part code for $x$, one part describing a Turing machine,
or {\em model}, for the {\em regular} aspects of $x$
and the second part describing
the {\em irregular} aspects of $x$ in the form
of a program to be interpreted by $T$, has
profound applications.

The ``right model'' is
a Turing machine $T$ among the ones that
halt for all inputs, a restriction that is justified later,
and reach the minimum description length in (\ref{eq.kcmdl}).
This $T$ embodies the amount of useful information
contained in $x$. It remains to decide
which such $T$ to select among the ones that satisfy
the requirement. Following Occam's Razor
we opt here for the shortest one---a formal justification
for this choice is given in \cite{ViLi99}. The main problem 
with our approach is how to properly define
a shortest program $x^*$ for $x$ that divides into parts $x^*=pq$ such
that $p$ represents an appropriate $T$. 

\subsection{Symmetry of Information}
The following central notions are used in this paper.
The {\em information in $x$ about $y$} is $I(x:y)= K(y)-K(y \mid x^*)$.
By the symmetry of information, a deep result of \cite{Ga74},
\begin{equation}\label{eq.soi}
K(x,y) \eqa K(x)+K(y \mid x^*) \eqa K(y)+K(x \mid y^*).
\end{equation}
Rewriting according to symmetry of information we see that
$I(x:y) \eqa I(y:x)$ and therefore we call the quantity
$I(x:y)$ the {\em mutual information} between $x$ and $y$. 

\section{Model Classes}
Instead of the model class of finite sets, or computable probability
density functions,
as in \cite{GTV}, in this work we focus on the most general form
of algorithmic model class: total recursive
functions. We define the different model classes
and summarize the central notions of ``randomness deficiency''
and ``typicality'' for the canonical finite set
models to obtain points of reference for the related notions
in the more general model classes.

\subsection{Set Models}
The  model class of {\em finite sets} consists of the set
of finite subsets $S \subseteq \{0,1\}^*$. The  {\em complexity
of the finite set} $S$ is
$K(S)$---the length (number of bits) of the
shortest binary program $p$ from which the reference universal
prefix machine $U$
computes a listing of the elements of $S$ and then
halts.
That is, if $S=\{x_1 , \ldots , x_{n} \}$, then
$U(p)= \langle x_1,\langle x_2, \ldots, \langle x_{n-1},x_n\rangle \ldots\rangle \rangle $.
The {\em conditional complexity} $K(x \mid S)$ of $x$ given $S$,
is the length (number of bits) in the
shortest binary program $p$ from which the reference universal
prefix machine $U$, given $S$ literally as auxiliary information,
computes $x$.
For every finite set $S  \subseteq \{0,1\}^*$ containing
$x$ we have
        \begin{equation}\label{eq57}
K(x  \mid  S)\lea\log|S|.
        \end{equation}
Indeed, consider the selfdelimiting code of $x$
consisting of its $\lceil\log|S|\rceil$ bit long index
of $x$ in the lexicographical ordering of $S$.
This code is called
\emph{data-to-model code}.
Its length quantifies the maximal ``typicality,'' or ``randomness,''
data (possibly different from $x$) can have with respect to this model.
The lack of typicality
of $x$ with respect to $S$
is measured by the amount by which $K(x \mid S)$
falls short of the length of the data-to-model code,
the {\em randomness deficiency} of $x$ in $S$, defined by
      \begin{equation}\label{eq:randomness-deficiency}
\delta (x  \mid  S) = \log |S| - K(x  \mid  S),
      \end{equation}
for $x \in S$, and $\infty$ otherwise.
Data $x$ is {\em typical with respect to a finite set} $S$,
if the randomness deficiency is small.
If the randomness deficiency is close to 0,
then there are no simple special properties that
single it out from the majority of elements in $S$.
This is not just terminology. Let $S \subseteq \{0,1\}^n$.
According to common viewpoints
in probability theory, each property represented by $S$ defines a large subset
of $S$ consisting of elements having that property, and, conversely,
each large subset of $S$ represents a property.
For probabilistic ensembles we take high probability subsets as
properties; the present case is uniform probability with finite support.
For some appropriate fixed constant
$c$, let us identify a property represented by
$S$ with a subset $S'$ of $S$ of cardinality
$|S'| > (1-1/c) |S|$.
  If $\delta (x  \mid  S)$ is close to 0, 
then $x$ satisfies (that is, is an element of) 
{\em all} properties (that is, sets) $S' \subseteq S$ 
of low Kolmogorov complexity
$K(S') = O(\log n)$.
The precise statements and quantifications are given in \cite{LiVi97,VeVi02},
and we do not repeat them here.

\subsection{Probability Models}
The model class of {\em computable probability
density functions} consists of the set
of functions $P: \{0,1\}^* \rightarrow [0,1]$ with
$\sum P(x) = 1$.
``Computable'' means here that there is a Turing machine $T_P$ that,
given $x$ and a positive rational  $\eps$,
computes $P(x)$ with precision $\eps$.
The (prefix-) complexity $K(P)$ of a
computable (possibly partial) function $P$ is defined by
$
K(P) = \min_i \{K(i): \mbox{\rm Turing machine } T_i
\; \; \mbox{\rm computes }
P \}.
$
                                                                                
\subsection{Function Models}
The model class of {\em total recursive
functions} consists of the set
of functions $f: \{0,1\}^* \rightarrow \{0,1\}^*$ such that
there is a Turing machine $T$ such that $T(i) < \infty$
and $f(i) = T(i)$, for every $i \in \{0,1\}^*$.
The (prefix-) complexity $K(f)$ of a
total recursive function $f$ is defined by
$
K(f) = \min_i \{K(i): \mbox{\rm Turing machine } T_i
\; \; \mbox{\rm computes }
f \}.
$
If $f^*$ is a shortest program for computing the function $f$
(if there is more than one of them then $f^*$ is the first one in
enumeration order), then $K(f)=l(f^*)$.
\begin{remark}
\rm
In the definitions of $K(P)$ and $K(f)$, the objects being
described are functions rather than finite binary strings.
To unify the approaches, we can 
consider a finite binary string $x$ as corresponding
to a function having value $x$ for argument 0.
Note that we can upper semi-compute $x^*$ given $x$,
but we cannot upper semi-compute $P^*$ given $P$ (as an oracle),
or $f^*$ given $f$ (again given as an oracle), since we should be able to 
verify agreement of a program for a function and an oracle for the 
target function, on all infinitely many arguments.
\end{remark}

\section{Typicality}
To explain typicality for general model classes, it is convenient
to use the distortion-rate
\cite{Sh48,Sh59} approach for individual data recently 
introduced in \cite{GV03,VV04}.
Modeling the data can be viewed as
encoding the data by a model: the data are source words
to be coded, and models are
code words for the data. As before, the set of possible data is 
${\cal D} = \{0,1\}^*$. Let ${\cal R}^+$ denote the set
of non-negative real numbers.
For every model class ${\cal M}$  (particular set of code words) 
we choose an appropriate 
recursive function 
$d: {\cal D} \times {\cal M} \rightarrow {\cal R}^+$ defining
the {\em distortion} $d(x,M)$ between data $x$ and model $M$.
\begin{remark}
\rm
The choice of distortion
function is a selection of which aspects of the data are relevant,
or meaningful, and
which aspects are irrelevant (noise).
We can think of the distortion as measuring how far the model 
falls short in representing the data. Distortion-rate theory
underpins the practice of lossy compression. 
For example, lossy compression of a sound file gives as ``model''
the compressed file where, among others, the very high and
very low inaudible frequencies have been suppressed. Thus,
the distortion function will penalize the deletion of the inaudible
frequencies but lightly because they are not relevant for the auditory
experience. 
\end{remark}
\begin{example}\label{ex.11}
\rm
Let us look at various model classes and distortion measures:

(i) The set of models are the finite sets of finite binary strings. 
Let $S \subseteq \{0,1\}^*$ and $|S| < \infty$.
We define $d(x,S) = \log |S|$ if $x \in S$, and $\infty$ otherwise.

(ii) The set of models are the computable probability density functions $P$
mapping $\{0,1\}^*$ to $[0,1]$. 
We define $d(x,S) =  \log 1/P(x)$ if $P(x) > 0$, and $\infty$ otherwise.

(iii) The set of models are the total recursive functions  $f$
mapping $\{0,1\}^*$ to ${\cal N}$. 
We define $d(x,f) = \min \{ l(d): f(d)=x\}$, and $\infty$ if
no such $d$ exists.
\end{example}
If ${\cal M}$ is a model class, then
we consider {\em distortion balls} of given
radius $r$ centered on $M \in {\cal M}$:
\[
B_M(r)= \{y: d(y,M) \leq r\}.
\]
This way, every model class and distortion measure can be treated 
similarly to the canonical finite set case, which, however is
especially simple in that the radius not variable.
That is, there is only one distortion ball  centered on a given finite set,
namely the one with radius equal to the log-cardinality of that finite set.
In fact, that distortion ball equals the finite set on which it is
centered.

Let ${\cal M}$ be a model class and $d$ a distortion measure.
Since in our definition the distortion is recursive,
given a model $M \in {\cal M}$ and diameter $r$, 
the elements in the distortion ball
of diameter $r$ can be recursively enumerated from the distortion function.
Giving the index of any element $x$ in that enumeration we can find the
element. Hence, $K(x|M,r) \lea \log |B_M(r)|$. On the other hand,
the vast majority of elements $y$ in the distortion ball have
complexity $K(y|M,r) \gea  \log |B_M(r)|$ since, for every constant $c$,
 there are only 
$2^{\log |B_M(r)|-c} - 1$ binary programs of length $ < \log |B_M(r)|-c$
available, and there are $|B_M(r)|$ elements to be described. 
We can now reason as in the similar case of finite set models.
With data $x$ and $r=d(x,M)$,
if $K(x|M,d(x,M))
\gea |B_M(d(x,M))|$, then $x$ belongs to every large majority of elements
(has the property represented by that majority)
of the distortion ball $|B_M(d(x,M))|$, provided that property is simple in the 
sense of having a description of low Kolmogorov complexity.   
\begin{definition}
\rm
the {\em randomness
deficiency} of $x$ with respect to model $M$ under distortion $d$
is defined as 
\[
\delta (x \mid M) = \log |B_M (d(x,M))| - K(x|M,d(x,M)). 
\]
Data $x$ is {\em typical} for model $M \in {\cal M}$ (and that model
``typical'' or ``best fitting'' for $x$) if 
\begin{equation}\label{eq.typical}
\delta (x \mid M)  \eqa 0.
\end{equation}
\end{definition}
If $x$ is typical for a model $M$, then the shortest way to effectively
describe $x$, given $M$, takes about as many bits as the 
descriptions of the great
majority of elements in
a recursive enumeration of the distortion ball. 
So there are no special simple properties that distinguish $x$ 
from the great majority of elements
in the distortion ball: they are all typical or random elements
in the distortion ball (that is, with respect to the contemplated model).
\begin{example}
\rm
Continuing Example~\ref{ex.11} by applying \eqref{eq.typical}
to different model classes:

(i) {\em Finite sets:}
 For finite set models $S$, clearly $K(x|S) \lea \log |S|$.
Together with \eqref{eq.typical} we have that $x$ is typical for $S$,
and $S$ best fits $x$, if the randomness deficiency 
according to \eqref{eq:randomness-deficiency} satisfies
$\delta(x|S) \eqa 0$. 

(ii) {\em Computable probability density functions:} 
Instead of the data-to-model code length $\log|S|$ for
finite set models, we consider the data-to-model code length
$\log 1/P(x)$ (the Shannon-Fano code). The value $\log 1/P(x)$
measures how likely $x$ is under the hypothesis $P$.
 For probability models $P$, 
define the conditional complexity
$K(x \mid P, \lceil  \log 1/P(x) \rceil )$ as follows. 
Say that a function
$A$ approximates $P$ if $|A(y,\eps)-P(y)|<\eps$
for every $y$ and every positive rational
$\eps$. Then $K(x \mid P , \lceil \log 1/P(x) \rceil)$ is defined as
the minimum length
of a program that, given $\lceil \log 1/P(x) \rceil$
and any function $A$ approximating $P$
as an oracle, prints $x$.

Clearly   
$K(x|P, \lceil \log 1/P(x) \rceil ) \lea \log 1/P(x)$. 
Together with \eqref{eq.typical}, we have that $x$ is typical for $P$,
and $P$ best fits $x$, if
$K(x|P, \lceil \log 1/P(x) \rceil) \gea \log |\{y:  \log 1/P(y) \leq 
 \log 1/P(x)\}|$. The right-hand side set condition is the same
as $P(y) \geq P(x)$, and there can be only $\leq 1/P(x)$ such $y$,
since otherwise the total probability exceeds 1. Therefore, 
the requirement, and hence typicality,
is implied by $K(x|P, \lceil \log 1/P(x) \rceil ) \gea \log 1/P(x)$.
Define  the randomness
deficiency by
$
\delta (x \mid P) =   \log 1/P(x) - K(x \mid P, \lceil \log 1/P(x) \rceil).
$
Altogether, a string $x$ is {\em typical for a distribution} $P$,
or $P$ is the {\em best fitting model} for $x$,
if $\delta (x \mid P) \eqa 0$. 
 
(iii) {\em Total Recursive Functions:} 
In place of $\log|S|$ for finite set models
we consider the data-to-model code length (actually, the distortion
$d(x,f)$ above)
$$\len xf=\min\{l(d):f(d)=x\}.$$
Define the conditional complexity
$K(x \mid f, \len xf )$ as
the minimum length
of a program that, given $\len xf$ and an oracle for  $f$,
prints $x$.

Clearly, $K(x|f, \len xf ) \lea \len xf$.
Together with \eqref{eq.typical}, we have that $x$ is typical for $f$,
and $f$ best fits $x$, if $K(x|f, \len xf ) \gea \log \{y: \len yf
 \leq  \len xf \}$. There are at most $(2^{\len xf +1} - 1)$-
many $y$ satisfying the set condition since
$\len yf \in \{0,1\}^*$.  Therefore,
the requirement, and hence typicality,
is implied by $K(x|f, \len xf ) \gea \len xf$.
Define  the  randomness
deficiency by
$
\delta (x \mid f) =   \len xf - K(x \mid f, \len xf ).
$
Altogether, a string $x$ is {\em typical for a total recursive
function} $f$, and $f$ is the {\em best fitting recursive function model}
for $x$ 
if $\delta (x \mid f) \eqa 0$, or written differently,
\begin{equation}\label{eq.typp}
K(x|f, \len xf ) \eqa \len xf.
\end{equation}
Note that since $\len xf$ is given as conditional information,
with $\len xf = l(d)$ and $f(d)=x$, the quantity $K(x|f, \len xf )$
represents the number of bits in a shortest 
{\em self-delimiting} description of $d$.  
\end{example}

\begin{remark}
\rm
We required $\len xf$ in the conditional in \eqref{eq.typp}.
This is the information about
the radius of the distortion ball centered on the model concerned.
Note that in the canonical finite set model case, as treated
in \cite{Ko74,GTV,VeVi02}, every model has a fixed radius which
is explicitly provided by the model itself. But in the 
more general model
classes of computable probability density functions, or
total recursive functions, models can have a variable radius. 
There are subclasses of the more general models that
have fixed radiuses (like the finite set models).

(i) In the computable probability density functions one can think of the
probabilities with a finite support, for example $P_n (x) = 1/2^n$
for $l(x)=n$, and $P(x)=0$ otherwise.

(ii) In the total recursive function case one can similarly think
of functions with finite support, for example $f_n (x) = \sum_{i=1}^n x_i$
for $x=x_1 \ldots x_n$, and $f_n(x)=0$ for $l(x) \neq n$.

The incorporation of te radius in the model will increase the
complexity of the model, and hence of the minimal sufficient statistic
below.
\end{remark}

\section{Sufficient Statistic}
A {\em statistic} is a function mapping the data to an element (model)
in the contemplated model class. With some sloppiness of terminology
we often call the function value (the model) also a statistic of the data.    
The most important concept in this paper is the sufficient statistic.
For an extensive discussion of this notion for specific model
classes see \cite{GTV,VeVi02}.
A statistic is called sufficient if the two-part description of
the data by way of the model and the data-to-model code is as 
concise as the shortest one-part description of $x$.
Consider a model class ${\cal M}$.
\begin{definition}
A model $M \in {\cal M}$ is a {\em sufficient statistic} for $x$ if
\begin{equation}\label{eq.ssm}
K(M, d(x,M))+ \log |B_M(d(x,M))| \eqa K(x).
\end{equation}
\end{definition}

\begin{lemma}\label{lem.V2}
If $M$ is a sufficient statistic for $x$, then 
$K(x \mid M, d(x,M) \eqa  \log |B_M(d(x,M))|$, that is,
$x$ is typical for $M$.
\end{lemma}
\begin{proof}
We can rewrite
$K(x) \lea K(x,M,d(x,M)) \lea K(M,d(x,M))+K(x|M,d(x,M))
\lea K(M, d(x,M))+ \log |B_M(d(x,M))| \eqa K(x)$.
The first three inequalities are straightforward and
the last equality is by the assumption of sufficiency.
Altogether, the first sum equals the second sum, which implies the lemma.
\end{proof}
Thus, if $M$ is a sufficient statistic for $x$, then $x$ is a typical element
for $M$, and $M$ is the best fitting model for $x$.
Note that the converse implication,  ``typicality'' implies
``sufficiency,'' is not valid. Sufficiency is a special type
of typicality, where the model does not add significant
information to the data, since the preceding proof shows 
$K(x) \eqa K(x,M,d(x,M))$. Using the symmetry of information \eqref{eq.soi}
this shows that 
\begin{equation}\label{eq.pcondx}
K(M,d(x,M) \mid x ) \eqa K(M \mid x) \eqa 0. 
\end{equation}
This means that:

(i) A sufficient statistic  $M$ is determined by the data in the sense
that we need only an $O(1)$-bit program, possibly depending on
the data itself, to compute the model
from the data.    

(ii) For each model class and distortion there is a universal constant $c$
such that for every data item $x$ there are at most $c$ sufficient
statistics.

\begin{example}
\rm
{\em Finite sets:} 
For the model class of finite sets, a set $S$ is a sufficient statistic
for data $x$ if
\[
K(S)+ \log |S| \eqa K(x).
\]

{\em Computable probability density functions:}
For the model class of computable probability density functions, 
a function $P$ is a sufficient statistic
for data $x$ if
\[
K(P)  + \log 1/P(x) \eqa 0.
\]
\end{example}
\begin{definition}
\rm
For the model class of 
{\em total recursive functions}, a function $f$ is a 
{\em sufficient statistic} for data $x$
if 
\begin{equation}\label{eq.ss}
K(x) \eqa K(f)  + \len xf .
\end{equation}
\end{definition}
Following the above discussion, the meaningful information in $x$
is represented by $f$ (the model) in $K(f)$ bits, and the
meaningless information in $x$ is represented by $d$ (the noise in
the data) with $f(d)=x$ in $l(d) = \len xf$ bits. Note that 
$l(d) \eqa K(d) \eqa K(d|f^*)$, 
since the two-part
code $(f^*,d)$ for $x$
cannot be shorter than the shortest one-part code of $K(x)$ bits,
and therefore the $d$-part must already be maximally compressed. 
By Lemma~\ref{lem.V2},  $\len xf \eqa   K(x \mid f^* , \len xf)$,
$x$ is typical for $f$,
and hence $K(x) \eqa K(f)  + K(x \mid f^* , \len xf)$.

\section{Minimal Sufficient Statistic}

\begin{definition}
\rm
Consider the model class of total recursive functions.
A {\em minimal sufficient statistic} for data $x$ is a sufficient statistic
\eqref{eq.ss}
for $x$ of minimal prefix complexity. Its length is
known as the {\em sophistication} of $x$, and  is defined by
$\soph (x) = \min \{ K(f): K(f)+ \len xf \eqa K(x) \}$.
\end{definition}

Recall that the {\em reference} universal prefix Turing machine 
$U$ was chosen such
that $U(T,d)=T(d)$ for all $T$ and $d$.
Looking at it slightly more from a programming point of view,
we can define
a pair $(T,d)$ to be a {\em description} of a finite
string $x$, if $U(T,d)$ prints $x$ and $T$ is a Turing machine
computing a function $f$ so 
that $f(d)=x$.
For the notion of minimal sufficient statistic
to be nontrivial, it should be impossible
to always shift, if $f(d)=x$ and $K(f)+\len xf \eqa K(x)$ with 
$K(f) \not\eqa 0$, always information 
information from $f$ to $d$ and write, 
for example, $f'(d')=x$ with $K(f')+ \len x{f'} \eqa K(x)$
with $K(f') \eqa 0$.
If the model class contains a fixed universal model that can 
mimic all other models, then we can always shift all model
information to the data-to-(universal) model code. Note that this
problem doesn't arise in common statistical model classes: these
do not contain universal models in the algorithmic sense.
First we show that the partial recursive
recursive function model class, because it contains a universal element,
does not allow a straightforward nontrivial
 division into meaningful and meaningless
information.

\begin{lemma}
Assume for the moment that we allow all partial recursive programs 
as statistic.
Then, the sophistication of all data $x$ is $\eqa 0$.
\end{lemma}

\begin{proof}
Let the index of $U$ (the reference universal prefix Turing machine)
in the standard enumeration $T_1,T_2, \ldots$
of prefix Turing machines be $u$. Let $T_f$ be a Turing machine
computing $f$. Suppose
that $U(T_f,d)=x$. Then, also
$U(u,\langle T_f,d\rangle)=U(T_f,d) = x$.
\end{proof}

\begin{remark}
\rm
This shows that unrestricted partial recursive statistics are 
uninteresting. Naively, this could leave the impression that
the separation of the regular and the random part of the data
is not as objective as the whole approach lets us hope for.
If we consider complexities of the minimal sufficient statistics
in model classes of increasing power:
finite sets, computable probability distributions, total recursive
functions, partial recursive functions, then the complexities
appear to become smaller all the time eventually reaching zero.
It would seem that the universality of Kolmogorov
complexity, based on the notion of partial recursive functions,
would suggest a similar universal notion of sufficient statistic
based on partial recursive functions. But in this case the very
universality trivializes the resulting definition: because 
partial recursive functions contain a particular universal element that
can simulate all the others, this implies that the universal
partial recursive function is a universal model for all data, and 
the data-to-model code incorporates all information in the data.
Thus, if a model class contains a universal model that can simulate
all other models, then this model class is not suitable for
defining two-part codes consisting of meaningful information and
accidental information. 
It turns out that the key to nontrivial separation is the
requirement that the program witnessing the sophistication be {\em total}.
That the resulting separation is non-trivial is evidenced by the fact,
shown below,
that the amount of meaningful information in the data does not
change by more than a logarithmic additive term under change of model
classes among finite set models, computable probability models, and
total recursive function models. That is, very different model classes
all result in the same amount of meaningful information in the data,
up to negligible differences. 
So if deterioration
occurs in widening model classes it occurs all at once by having a 
universal element in the model class.
\end{remark}

Apart from triviality, a class of
statistics can also possibly
be vacuous by having the length of the minimal sufficient
statistic exceed $K(x)$.
Our first task is to determine whether the definition
is non-vacuous.
We will distinguish
sophistication in different description modes:


\begin{lemma}[Existence]\label{lem.exists}
For every finite binary string $x$, the sophistication 
satisfies $\soph (x) \lea K(x)$.
\end{lemma}
\begin{proof}
By definition of the prefix complexity there is a program $x^*$
of length $l(x^*)=K(x)$ such that $U(x^*, \epsilon) = x$.
This program $x^*$ can be partial. But we can define another program
$x^*_s = sx^*$ where $s$ is a program of a constant number of bits that
tells the following program to ignore its actual input and compute 
as if its input were $\epsilon$. Clearly, $x^*_s$ is total and 
is a sufficient statistic of the total recursive function type,
that is,
$\soph (x)\leq l(x^*_s) \lea l(x^*) = K(x)$.
\end{proof}
The previous lemma gives an upper bound on the sophistication.
This still leaves the possibility that the sophistication is always $\eqa 0$,
for example in the most liberal case of unrestricted totality.
But this turns out to be impossible.

\begin{theorem} \label{h-sophI}
(i) For every $x$, if a sufficient statistic $f$
satisfies $K(\len xf |f^*) \eqa 0$, then
 $K(f) \gea K(K(x))$  and
$\len xf \lea K(x) - K(K(x))$.

(ii)
For $x$ as a variable running through a sequence of finite binary strings
of increasing length, we have
\begin{equation}\label{eq.liminf}
\liminf_{l(x) \rightarrow \infty} \soph (x) \eqa 0.
\end{equation}

(iii) 
For every $n$, 
there exists an $x$ of length $n$, 
such that every sufficient statistic $f$ for $x$
that satisfies $K(\len xf |f^*) \eqa 0$ has
$K(f) \gea n$.

(iv)
For every $n$ there exists an $x$
of length $n$
such that $\soph(x)\gea n - \log n - 2 \log \log n$.
\end{theorem}

\begin{proof}
(i) If $f$ is a sufficient statistic for $x$, then
\begin{equation}\label{eq.pd}
 K(x) \eqa K(f)+K(d \mid f^*) \eqa K(f)+ \len xf.
\end{equation}
Since $K(\len xf |f^*) \eqa 0$, given an $O(1)$ bit program $q$ we
can retrieve both $\len xf$ and
and also $K(f) = l(f^*)$ from $f^*$. Therefore, 
we can retrieve $K(x) \eqa K(f)+ \len xf$ from $qf^*$.
That shows that $K(K(x)) \lea K(f)$.
This proves both the first statement, and the second statement follows 
by (\ref{eq.pd}).

(ii)
An example of very unsophisticated
strings are the individually random strings with high complexity:
$x$ of length $l(x) = n$ with complexity $K(x) \eqa n + K(n)$.
Then, the {\em identity} program $\iota$ with $\iota(d)=d$ for all $d$ is total,
has complexity $K(\iota) \eqa 0$, and satisfies $K(x) \eqa K(\iota) + l(x^*)$.  
Hence, $\iota$ witnesses that $\soph (x) \eqa 0$. This shows
(\ref{eq.liminf}).

(iii) 
 Consider the set $S^m = \{y: K(y) \leq m \}$. By \cite{GTV}
we have $\log |S^m| \eqa m - K(m)$. Let $m \leq n$.
Since there are $2^n$
strings of length $n$, there are strings of length $n$
not in $S^m$. Let $x$ be any such string, and denote $k=K(x)$.
Then, by construction $k > m$ and by definition $k \lea n + K(n)$.
Let $f$ be a sufficient statistic for $x$.
Then, $K(f)+ \len xf \eqa  k$. By assumption, there is an $O(1)$-bit
program $q$ such that $U(qf^*)= \len xf$.
Let $d$ witness $\len xf$ by $f(d)=x$ with $l(d) = \len xf$. Define
the set $D= \{0,1\}^{\len xf}$.
Clearly, $d \in D$.
Since $x$ can be retrieved from $f$ and the lexicographical index
of $d$ in $D$, and $\log |D| = \len xf$, we have $K(f)+ \log |D| \eqa k$. 
Since we can obtain $D$ from $qf^*$ we have $K(D) \lea K(f)$.
On the other hand, since we can retrieve $x$ from $D$ and the index
of $d$ in $D$, we must have $K(D)+\log |D| \gea k$, which implies
$K(D) \gea K(f)$. Altogether, therefore, $K(D) \eqa K(f)$.

We now show that we can choose $x$ so that $K(D) \gea n$,
and therefore $K(f) \gea n$.
For every length $n$, there exists a $z$ of complexity
$K(z \mid n) \lea n$ such that a
minimal sufficient finite set statistic $S$ for $z$
has complexity at least $K(S \mid n) \gea n$, by Theorem IV.2
of \cite{GTV}. Since $\{z\}$ is trivially a sufficient
statistic for $z$, it follows $K(z \mid n) \eqa K(S \mid n) \eqa n$. 
This implies $K(z),K(S) \gea n$. 
Therefore, we can choose $m = n - c_2$ for a large
enough constant $c_2$
so as to ensure that 
$z \not\in S^m$. Consequently, we can choose $x$ above as such a $z$.
Since every finite set sufficient statistic for $x$
has complexity at least that of an 
finite set minimal sufficient statistic for $x$, 
it follows that $K(D) \gea n$. Therefore,
$K(f) \gea n$, which was what we had to prove.  

(iv)
In the proof of (i) we used $K( \len xf | f^*) \eqa 0$.
Without using this assumption, the corresponding argument
yields $k \lea K(f)+K( \len xf ))+ \log |D|$.
We also have $K(f) + \len xf \lea k$ and $l(d) \eqa \log |D|$.
Since we can retrieve $x$ from $D$ and its index in $D$, the same
argument as above shows $|K(f) - K(D)| \lea K( \len xf )$, and
still following the argument above, $K(f) \gea n - K( \len xf )$.
Since $\len xf  \lea n$ we have $K(\len xf ) \lea \log n + 2 \log \log n$.
This proves the statement.
\end{proof}

The useful (\ref{eq.pcondx}) states  that there is a constant,
such that for every $x$ there are
at most that constant many sufficient statistics for $x$,
and there is a constant
length program (possibly depending on $x$),
 that generates all of them from $x^*$. In fact,
there is a slightly stronger statement from which this follows:

\begin{lemma}
There is a universal constant $c$,
such that for every $x$,
the number of $f^*d$ such that $f(d) = x$ and $K(f)+l(d) \eqa K(x)$,
is bounded above by $c$.
\end{lemma} 

\begin{proof}
Let the prefix Turing machine $T_f$ compute $f$.
Since $U(T_f,d)=x$ and $K(T_f)+l(d) \eqa K(x)$, 
the combination $f^* d$
(with self-delimiting $f^*$) is a shortest prefix program for $x$.
From \cite{LiVi97}, Exercise 3.3.7 item (b) on p. 205, it follows
that the number of shortest prefix programs is upper bounded
by a universal constant. 
\end{proof}

\section{Relation Between Sufficient Statistic for Different Model Classes}
Previous work studied sufficiency for
finite set models, and computable probability mass functions models,
\cite{GTV}. 
The most general models that are still meaningful
are total recursive functions as studied here.
We show that there are corresponding, almost equivalent,
sufficient statistics in all model classes. 

\begin{lemma}\label{lem.explimpl}

(i) If $S$ is a sufficient statistic of $x$ (finite set type),
then there is a corresponding
sufficient statistic $P$ of $x$ (probability mass function type) such that
$K(P) \eqa K(S)$, $ \log 1/P(x) \eqa \log |S|$, and $K(P \mid x^*) \eqa 0$.   

(ii) If $P$ is a sufficient 
statistic of $x$ of the computable total probability density function type,
then there is a corresponding
sufficient statistic $f$ of $x$ of the total recursive function type 
such that
$K(f) \eqa K(P)$, 
$\len xf \eqa  \log 1/ P(x)$, and $K(f \mid x^*) \eqa 0$.   
\end{lemma}

\begin{proof}
(i) By assumption, $S$ is a finite set such that $x \in S$
and $K(x) \eqa K(S)+ \log |S|$.
Define the probability distribution $P(y) = 1/|S|$ for $y \in S$
and $P(y)=0$ otherwise. Since $S$ is finite, $P$ is computable.
Since $K(S) \eqa K(P)$, and $\log |S| = \lceil \log 1/P(x) \rceil$,
we have $K(x) \eqa K(P) + \log 1/P(x)$.
Since $P$ is a computable probability mass function we have
$K(x \mid P^*) \lea  \log 1/P(x)$, by the standard Shannon-Fano
code construction \cite{CT91} that assigns a code word of length
$ \log 1/ P(x)$ to $x$. Since by (\ref{eq.soi}) we have $K(x) \lea K(x,P) \eqa
K(P)+ K(x \mid P^*)$ it follows that $ \log 1/P(x) \lea K(x \mid P^*)$.
Hence, $K(x \mid P^*) \eqa  \log 1/P(x)$. 
Therefore, by (\ref{eq.soi}), 
$K(x,P) \eqa K(x)$ and, by  rewriting $K(x,P)$
in the other way according to (\ref{eq.soi}), $K(P \mid x^*) \eqa 0$.

(ii) By assumption, $P$ is a
computable probability
density function with  $P(x) > 0$ and
$K(x) \eqa K(P)  + \log 1/ P(x)$.
The witness of this equality is a shortest program $P^*$ for
$P$ and a code word $s_x$ for $x$ according to the standard
Shannon-Fano code, \cite{CT91}, with $l(s_x) \eqa  \log 1/P(x)$.
Given $P$, we can reconstruct $x$ from $s_x$ by a fixed standard
algorithm.
Define the recursive function $f$ from $P$ such that
$f(s_x) = x$. In fact, from $P^*$ this only requires
a constant length program $q$, so that $T_f=qP^*$ is a program
that computes $f$ in the sense that $U(T_f,d)=f(d)$ for all $d$. 
Similarly, $P$ can be retrieved from $f$. Hence, $K(f) \eqa K(P)$ and
$K(x) \eqa K(f) + l(s_x)$. 
That is, $f$ is a sufficient
statistic for $x$.
Also, $f$ is a total recursive function.
Since $f(s_x)=x$ we have $K(x \mid f^*) \lea l(s_x)$,
and $K(x \mid f^*) \lea l(s_x)$.
This shows that $K(x) \gea K(f) +K(x \mid f^*)$, and since $x$
can by definition
be reconstructed from $f^*$ and a program of length $K(x \mid f^*)$,
it follows that equality must hold.
Consequently, $l(s_x) \eqa K(x \mid f^*)$, and hence,
by (\ref{eq.soi}),  $K(x,f) \eqa K(x)$ and $K(f \mid x^*) \eqa 0$.
\end{proof}

We have now shown that a sufficient statistic in
a less general model class corresponds directly to a sufficient
statistic in the next more general model class. We now show that,
with a negligible error term, a sufficient statistic in the most
general model class of total recursive functions has a
 directly corresponding sufficient
statistic in the least general finite set model class. That is,
up to negligible error terms, a sufficient statistic in any of the model classes
has a direct representative in any of the other model classes.

\begin{lemma}
Let $x$ be a string of length $n$, and
$f$ be a total recursive function sufficient statistic
for $x$. Then, there is a finite set $S \ni x$  such that
$K(S)+\log |S| \eqa K(x)+O(\log n)$. 
\end{lemma}

\begin{proof}
By assumption there is an $O(1)$-bit program $q$ such that
$U(qf^*)= \len xf$.
For each $y \in \{0,1\}^*$, 
let $i_y = \min \{i : f(i)=y\}$.
Define $S=\{y:f(i_y)=y, l(i_y) \lea \len xf \}$. 
We can compute $S$ by computation of $f(i)$,
on all arguments $i$ of at most $l(i) \leq \len xf $ bits, 
since by assumption $f$ is total. This shows
$K(S) \lea K(f, \len xf) \lea K(f)+ K( \len xf )$.
Since $\len xf \lea K(x)$, we have $\len xf \lea l(x)=n$.
Moreover, $\log |S| \eqa \len xf$.
Since $x \in S$, 
$K(x) \lea K(S)+\log|S| \lea K(x)+O(\log n)$, where we use the sufficiency
of $f$ to obtain the last inequality.
\end{proof}

\section{Algorithmic Properties}

We investigate the recursion properties of the sophistication
function. 
In \cite{Ga74}, G\'acs gave 
an important and deep result 
(\ref{eq.gacs}) below, that quantifies 
the uncomputability of $K(x)$ (the bare uncomputability can be established
in a much simpler fashion). 
For every length $n$ there is an $x$ of length $n$ such that:
\begin{equation}\label{eq.gacs}
\log n - \log \log n \lea K(K(x) \mid x ) \lea \log n .
\end{equation}
Note that the right-hand side holds for every $x$ by
the simple argument that $K(x) \leq n + 2 \log n$ and hence
$K(K(x)) \lea \log n$. But there are $x$'s such that the length
of the shortest program to compute $K(x)$ almost
reaches this upper bound, even if the full information
about $x$ is provided. 
It is natural to suppose that 
the sophistication function is not recursive either. 
The following lemma's suggest that the complexity
function is more uncomputable than the sophistication.

\begin{theorem}
The function $\soph$
is not recursive.
\end{theorem}

\begin{proof}
Given $n$,
let $x_0$
be the least $x$ such that $\soph (x) > n - 2 \log n$.
By Theorem~\ref{h-sophI}
we know that there exist $x$ such that 
$\soph (x) \rightarrow \infty$ for $x \rightarrow \infty$,
hence $x_0$ exists. Assume by way of contradiction that the
sophistication function is computable. Then, we can find $x_0$, given $n$,
by simply computing the successive values of the function. But then
$K(x_0) \lea K(n)$, while by Lemma~\ref{lem.exists}
 $K(x_0) \gea \soph (x_0)$ and by 
assumption $\soph (x_0) > n - 2 \log n$,
which is impossible.
\end{proof}

The {\em halting sequence} $\chi = \chi_1 \chi_2 \ldots$
is the infinite binary characteristic sequence of the halting problem,
defined by $\chi_i = 1$
if the reference universal prefix Turing machine $U$
halts on the $i$th input: $U(i) < \infty$, and 0 otherwise.

\begin{lemma}\label{lem.compKs}
Let $f^*$ be a total recursive function sufficient statistic of $x$.

(i) We can compute $K(x)$ from $f^*$ and $x$, up to fixed
constant precision, which implies that
$K(K(x) \mid  f^*,x ) \eqa 0$.

(ii) If also $K ( \len xf | f^*) \eqa 0$, then 
we can compute $K(x)$ from $f^*$,  up to fixed
constant precision, which implies that
$K(K(x) \mid  f^* ) \eqa 0$.
\end{lemma}
\begin{proof}
(i) 
Since $f$ is total, we can
run $f(e)$ on all strings $e$ in lexicographical length-increasing
order. Since $f$ is total we will find a shortest string $e_0$
such that $f(e_0)=x$. Set $\len xf = l(e_0)$. Since $l(f^*)=K(f)$,
and by assumption,
$K(f)+  \len xf \eqa K(x)$, 
we now can compute $\eqa K(x)$.

(ii) follows from item (i).
\end{proof}

\begin{theorem}
Given an oracle that on query $x$ answers with a 
sufficient statistic $f^*$ of $x$ and a $c_x \eqa 0$ as required below.
Then, we can compute the Kolmogorov complexity 
function $K$
and the halting sequence $\chi$.
\end{theorem}

\begin{proof}
By Lemma~\ref{lem.compKs} we can compute the function $K(x)$,
up to fixed constant precision,
given the oracle (without the value $c_x$)
 in the statement of the theorem. Let $c_x$ in
the statement of the theorem be the difference
between the computed value and the actual value of $K(x)$. In
\cite{LiVi97}, Exercise 2.2.7 on p. 175, it is shown that
if we can solve the halting problem for plain Turing machines, 
then we can compute
the (plain) Kolmogorov complexity, and {\em vice versa}.
The same holds for the halting problem for prefix Turing machines
and the prefix Turing complexity.
This proves the theorem.
\end{proof}

\begin{lemma}\label{lem.chisoph}
There is a constant $c$, such that for every $x$
there is a program (possibly depending on $x$)
of at most $c$ bits that computes $\soph (x)$
and the
witness program $f$ from $x,K(x)$.
That is, $K(f \mid x,K(x)) \eqa 0$.
With some abuse of notation we can express this as
$K(\soph \mid K) \eqa 0$.
\end{lemma}

\begin{proof}
By definition of sufficient statistic $f^*$, 
we have $K(f)+ \len xf \eqa K(x)$. 
By (\ref{eq.pcondx}) 
the number of
sufficient statistics 
for $x$ is bounded by an independent constant, 
and we can generate all of them from $x$
by a $\eqa 0$ length program (possibly depending on $x$). Then,
we can simply determine the least length of a sufficient
statistic, which is
$\soph (x)$.
\end{proof}

There is a subtlety here: Lemma~\ref{lem.chisoph}
is nonuniform. While for every $x$ we only require a fixed number
of bits to compute the sophistication from $x,K(x)$, the  result is nonuniform
in the sense that these bits may depend on $x$. Given a 
program, how do we
verify if it is the correct one? Trying all programs
of length up to a known upper bound, we don't know
if they halt or if they halt they halt with the correct
answer. The question arising
is if there is a single program that computes the sopistication  
and its witness program for all $x$. In \cite{VeVi02}
this much more difficult question is answered in a strong negative sense: 
there is no algorithm that for every $x$, given $x,K(x)$,
 approximates the sophistication
of $x$ to within precision $l(x)/(10 \log l(x))$.

\begin{theorem}
For every $x$ of length $n$, and $f^*$ the program
that witnesses the sophistication of $x$,
we have $K(f^* \mid x) \lea \log n$.
For every length $n$, there are strings $x$ of length $n$,
such that $K(f^* \mid x) \gea \log n - \log \log n$. 
\end{theorem}

\begin{proof}
Let $f^*$ witness the $\soph(x)$: That is, 
$K(f)+ \len xf \eqa K(x)$, 
and $l(f^*) = \soph (x)$.
Using the conditional
version of (\ref{eq.soi}), see \cite{GTV},
we find that
$K(K(x),f^* \mid x)$
\begin{align*}
& \eqa K(K(x) \mid x) + K(f^* \mid K(x), K(K(x) \mid x),x)
\\ & \eqa K(f^* \mid x) + K(K(x) \mid f^*, K(f^* \mid x),x).
\end{align*}
In Lemma~\ref{lem.compKs}, item (i), we show $K(K(x) \mid x,f^*) \eqa 0$,
hence also $K(K(x) \mid f^*, K(f^* \mid x),x) \eqa 0$.
By Lemma~\ref{lem.chisoph},
$K(f^* \mid K(x),x) \eqa 0$, hence also
$K(f^* \mid K(x), K(K(x) \mid x),x) \eqa 0$. Substitution
of the constant terms in the displayed equation shows
\begin{equation}\label{eq.pK}
K(K(x),f^* \mid x) \eqa K(f^* \mid x) \eqa K(K(x) \mid x) \eqa K(x^* \mid x).
\end{equation}

This shows that the shortest program to retrieve $f^*$
from $x$ is
essentially the same program as to retrieve $x^*$ from $x$ or
$K(x)$ from $x$.
Using \eqref{eq.gacs}, this shows that
\[
\log l(x) \gea \limsup_{l(x) \rightarrow \infty} K(f^* \mid x) \gea
\log l(x) - \log \log l(x).
\]
Since $f^*$ is the witness program for $l(f^*) = \soph (x)$,
we have  $l(f^*) \eqa K(f^*) \gea K(f^* \mid x)$.
\end{proof}

\begin{definition}
\rm
A function $f$ from the rational numbers to the real numbers
 is {\em upper semicomputable} if 
there is a recursive function $H(x,t)$ such that $H(x,t+1) \leq H(x,t)$
and $\lim_{t \rightarrow \infty} H (x,t)=f(x)$. Here we interprete
the total recursive function $H(\langle x,t \rangle) = \langle p,q\rangle$
as a function from pairs of natural numbers to the rationals:
$H(x,t)=p/q$. If $f$ is upper semicomputable, 
then $-f$ is {\em lower semicomputable}.
If $f$ is both upper-a and lower semicomputable, then it is {\em computable}.
\end{definition}
Recursive functions are computable functions over the natural numbers.
Since $K(\cdot)$ is upper semicomputable, \cite{LiVi97}, and from $K(\cdot)$
we can compute $\soph (x)$, we have the following:

\begin{lemma}
(i) The function $\soph (x)$ 
is not computable
to any significant precision.

(ii)  Given an initial segment
of length $2^{2l(x)}$ of the halting sequence $\chi = \chi_1 \chi_2 \ldots$, 
we can compute $\soph (x)$
from $x$.
That is, 
$K(\soph (x) \mid x, \chi_1 \ldots \chi_{2^{2l(x)}} ) \eqa 0$.
\end{lemma}

\begin{proof}
(i) 
The fact that $\soph (x)$ is not computable to any significant precision
is shown in \cite{VeVi02}.

(ii) We can run $U(p,d)$ for all (program, argument)
pairs such that $l(p)+l(d) \leq 2 l(x)$. (Not $l(x)$ since we
are dealing with self-delimiting programs.)
 If we know  the initial segment of $\chi$,
as in the statement of the theorem, then we 
know which (program, argument) pairs halt,
and
we can simply compute the minimal value of $l(p)+l(d)$
for these pairs.
\end{proof}

\section{Discussion}
``Sophistication'' is the algorithmic
version of ``minimal sufficient statistic'' for data $x$ in the model class
of total recursive functions. However, the full stochastic properties
of the data can only be understood by considering the Kolmogorov structure
function $\lambda_x (\alpha)$ (mentioned earlier)
that gives the length of the shortest two-part code of $x$
as a function of the maximal complexity $\alpha$ of the total function
supplying the model part of the code. This function has value
about $l(x)$ for $\alpha$ close to 0, 
is nonincreasing, and drops to the
line $K(x)$ at complexity $\alpha_0 = \soph(x)$, after which it
remains constant, $\lambda_x(\alpha)=K(x)$ for $\alpha \geq \alpha_0$,
everything up to a logarithmic addive term. 
A comprehensive analysis, including many more
algorithmic properties than are analyzed here, has been given in \cite{VeVi02}
for the model class of finite sets containing $x$, but it is shown there
that all results extend to the model class of computable probability
distributions and the model class of total recursive functions,
up to an additive logarithmic term.

%
%
%
\section*{Acknowledgment}
The author thanks Luis Antunes, Lance Fortnow, Kolya Vereshchagin,
 and the referees for their
comments.

\section*{Biography}

{\sc Paul M.B. Vit\'anyi} is a Fellow
of the Center for Mathematics and Computer Science (CWI)
in Amsterdam and is Professor of Computer Science
at the University of Amsterdam.  He serves on the editorial boards
of Distributed Computing (until 2003), Information Processing Letters,
Theory of Computing Systems, Parallel Processing Letters,
International journal of Foundations of Computer Science,
Journal of Computer and Systems Sciences (guest editor),
and elsewhere. He has worked on cellular automata,
computational complexity, distributed and parallel computing,
machine learning and prediction, physics of computation,
Kolmogorov complexity, quantum computing. Together with Ming Li
they pioneered applications of Kolmogorov complexity
and co-authored ``An Introduction to Kolmogorov Complexity
and its Applications,'' Springer-Verlag, New York, 1993 (2nd Edition 1997),
parts of which have been translated into Chinese,  Russian and Japanese.

\end{document}